\documentclass[]{spie}  
\newcommand{\ElectronCreation}{a_{\scriptsize{c \mbox{{\bfseries k}}}}^{\dag}}
\newcommand{\HoleCreation}{b_{\scriptsize{v \mbox{{\bfseries k}}}}^{\dag}}
\newcommand{\wv}{\mbox{{\bfseries k}}}
\newcommand{\omegac}{\omega_{\scriptsize{c}} \! \left( \mbox{{\bfseries k}} \right) }
\newcommand{\omegav}{\omega_{\scriptsize{v}} \! \left( \mbox{{\bfseries k}} \right) }
\newcommand{\omegacv}{\omega_{\scriptsize{cv}} \! \left( \mbox{{\bfseries k}} \right) }

\usepackage{amsmath,amsfonts,amssymb}
\usepackage{graphicx}
\usepackage[colorlinks=true, allcolors=blue]{hyperref}

\title{Ultrafast orbital manipulation and Mott physics in multi-band correlated materials}

\author[a,b,c]{Andrea Ronchi}
\author[b]{Paolo Franceschini}
\author[d]{Laura Fanfarillo}
\author[a]{P\'ia Homm}
\author[a]{Mariela Menghini}
\author[c]{Simone Peli}
\author[b,c]{Gabriele Ferrini}
\author[b,c]{Francesco Banfi}
\author[e]{Federico Cilento}
\author[f,g]{Andrea Damascelli}
\author[e,h,i]{Fulvio Parmigiani}
\author[a]{Jean-Pierre Locquet}
\author[d]{Michele Fabrizio}
\author[d]{Massimo Capone}
\author[b]{Claudio Giannetti}
\affil[a]{Department of Physics and Astronomy, KU Leuven, Celestijnenlaan 200D, 3001 Leuven, Belgium}
\affil[b]{Department of Mathematics and Physics, Universit\`a Cattolica del Sacro Cuore, Brescia I-25121, Italy}
\affil[c]{Interdisciplinary Laboratories for Advanced Materials Physics (ILAMP), Universit\`a Cattolica del Sacro Cuore, Brescia I-25121, Italy}
\affil[d]{Scuola Internazionale Superiore di Studi Avanzati (SISSA) and CNR-IOM Democritos National Simulation Center, Via Bonomea 265, 34136 Trieste, Italy}
\affil[e]{Elettra-Sincrotrone Trieste S.C.p.A., 34149 Basovizza, Italy}
\affil[f]{Department of Physics and Astronomy, University of British Columbia, Vancouver, BC V6T 1Z1, Canada}
\affil[g]{Quantum Matter Institute, University of British Columbia, Vancouver, BC V6T 1Z4, Canada}
\affil[h]{Dipartimento di Fisica, Universit\`a degli Studi di Trieste, 34127 Trieste, Italy}
\affil[i]{International Faculty, University of Cologne, Albertus-Magnus-Platz, 50923 Cologne, Germany}

\authorinfo{Further author information: (Send correspondence to M.C. and C.G.)\\
M.C.: E-mail: capone@sissa.it
\\C.G.: E-mail: claudio.giannetti@unicatt.it, Telephone: +39 030 2406709, \\Webpage: \href{http://docenti.unicatt.it/eng/claudio_giannetti/}{C. Giannetti webpage}, lab: \href{http://centridiricerca.unicatt.it/ilamp-ulysses-research-2022}{Ulysses@I-LAMP}}

\begin{document} 
\maketitle

\begin{abstract}
Multiorbital correlated materials are often on the verge of multiple electronic phases (metallic, insulating, superconducting, charge and orbitally ordered), which can be explored and controlled by small changes of the external parameters. The use of ultrashort light pulses as a mean to transiently modify the band population is leading to fundamentally new results. In this paper, we will review recent advances in the field and we will discuss the possibility of manipulating the \textit{orbital polarization} in correlated multi-band solid state systems. This technique can provide new understanding of the ground state properties of many interesting classes of quantum materials and offers a new tool to induce transient emergent properties with no counterpart at equilibrium. We will address: the discovery of high-energy Mottness in superconducting copper oxides and its impact on our understanding of the cuprate phase diagram; the instability of the Mott insulating phase in photoexcited vanadium oxides; the manipulation of orbital-selective correlations in iron-based superconductors; the pumping of local electronic excitons and the consequent transient effective quasiparticle cooling in alkali-doped fullerides. Finally, we will discuss a novel route to manipulate the \textit{orbital polarization} in a {\wv}-resolved fashion.  
\end{abstract}


\section{INTRODUCTION}
\label{sec:intro}  
   \begin{figure} [t]
   \begin{center}
   \begin{tabular}{c} 
   \includegraphics[height=10cm]{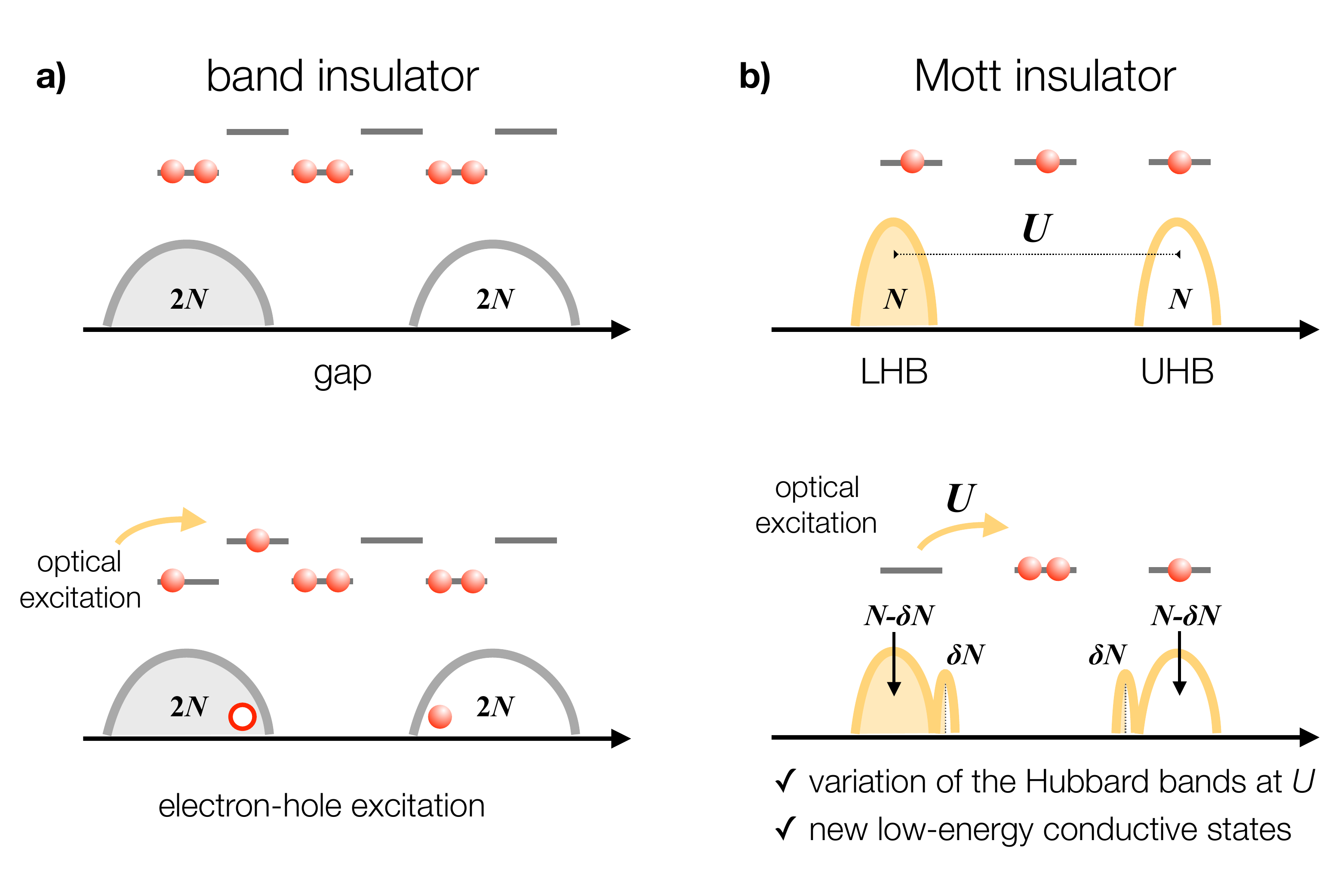}
   \end{tabular}
   \end{center}
   \caption[figure1] 
   { \label{figure1} 
Cartoon of the photoexcitation process in solids. \textbf{a)} Photoexcitation of a conventional band insulator: the electronic occupation is transiently modified within a rigid bandstructure. \textbf{b)} Photoexcitation of a Mott insulator: the multiplicity of the upper- (UHB) and lower-Hubbard bands (LHB) is modified by the optical excitation and is compensated by the emergence of new conductive states at low energies.}
   \end{figure} 
The intrinsic fragility of correlated materials is at the base of the recent ascent of techniques and protocols for achieving the complete optical control of their properties. The idea underlying this intense research activity is the use of light as an external stimulus to manipulate the electronic properties of materials in the vicinity of phase transitions or charge-spin-lattice instabilities. When scaled down to the ultrafast timescale, femtosecond light pulses can be used as a knob to control the band occupation and create transient electronic distributions with no counterpart at equilibrium \cite{Orenstein2012,Zhang2014,Giannetti2016,Gandolfi2017,Basov2017}.
Correlated materials constitute a natural playground for this non-equilibrium physics as a consequence of the intrinsic instability of their bandstructure, which has roots in the way the electronic bands are built from the atomic orbitals of the $N$ atoms forming the solid. In conventional band insulators, the formation of the insulating gap is the consequence of the lack of states in a specific energy range. In the simplest picture, corresponding to the two-band model, the completely filled valence band originates from the $N$ atomic orbitals which are fully occupied by two electrons/atom. Under moderate photoexcitation, the bandstructure remains rigid while electron-hole excitations are injected into the system, thus modifying the equilibrium charge distribution within the bands. In correlation-driven insulators, commonly named as Mott insulators, the gap opens already in a single-band picture, since it originates from the strong Coulomb repulsion, $U$, between two electron occupying the same orbital. When the value of $U$ overcomes the  \textit{bare} bandwidth, $W$, the electron motion is frozen, despite the fact that the $N$ atomic orbitals contain an odd number of electrons. This process leads to the formation of two characteristic bands separated by $U$ (Lower Hubbard Band, filled; Upper Hubbard Band, empty), whose multiplicity is related to the number $N$ of combinations that are available for getting a double occupancy on one of the sites. It is thus clear that any photo-induced electron-hole excitation, at the energy cost $U$, creates additional conductive states (corresponding to the holes and doublons which can move throughout the lattice at zero energy cost) and also modifies the multiplicity of the LHB and UHB. The basic consequence of the injection of $\delta N$  photoexcitations across the Mott gap is thus the emergence of $\delta N$ low-energy metallic-like states and the corresponding decrease of the spectral weight of electronic bands at the energy scale $U$. 
In the present manuscript we will review the impact of electronic correlations and Mott physics on the properties of multi-band transition-metal (TM) compounds, with particular focus on systems which develop superconductivity at low temperature. Building on the interplay between the band occupation and the electronic bandstructure introduced above, we will discuss the most promising strategies to control the electronic properties of these materials via selective light excitation. In many cases, the possibility to set up correlated systems with a non-equilibrium band occupation enables to tackle long-standing questions from a new viewpoint.

\section{ULTRAFAST ORBITAL MANIPULATION IN MULTI BAND MATERIALS}
  
\subsection{Mott physics in multi-band materials}
\begin{figure} [t]
   \begin{center}
   \begin{tabular}{c} 
   \includegraphics[height=5cm]{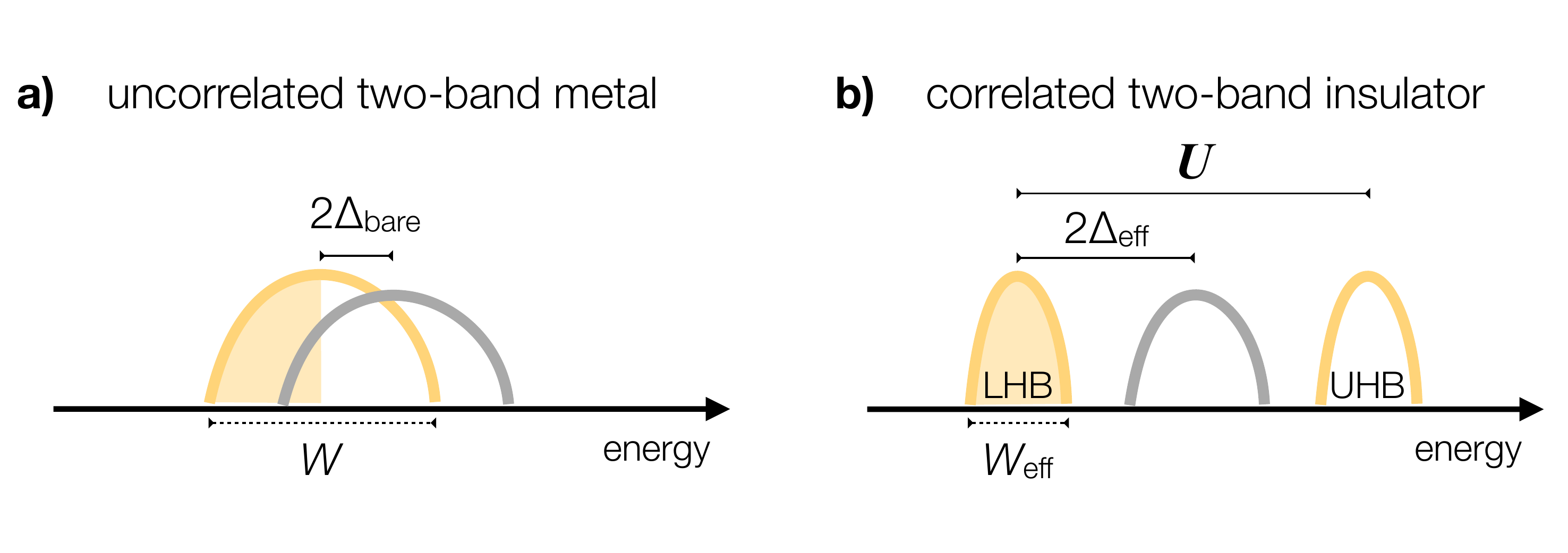}
   \end{tabular}
   \end{center}
   \caption[figure1] 
   { \label{fig:multiband} 
Cartoon of the bandstructure of a two-band solid. \textbf{a)} Uncorrelated two-band metal ($U$=0). The overlapping band are separated by a bare energy distance 2$\Delta_{\mathrm{bare}}$ smaller than the bare bandwidth $W$. \textbf{b)} Correlated two-band insulator with $U> W> 2\Delta_{\mathrm{bare}}$. The Coulomb repulsion simultaneously renormalizes the bandwidth and increases the effective gap as expressed by Eq. \ref{eff_gap}.}
   \end{figure}

Transition-metal (TM) compounds constitute one of the most interesting families of materials in which electronic correlations are at the heart of a rich variety of phenomena, ranging from high-temperature superconductivity to insulator-to-metal transitions and unconventional magnetism. The partial filling of the TM $d$ shells leads to a strong onsite Coulomb repulsion ($U\sim$1-10 eV) which determines some universal features, such as: 
i) the strong frustration of the electronic kinetic energy, that leads to the narrowing of the effective bandwidth ($W_{\mathrm{eff}}$) as compared to the bare one; ii) the emergence of robust short-range magnetic correlations, which often give rise to long-range antiferromagnetic phases. At the same time, the complex chemical composition of these compounds brings in the multi-band (orbital) physics as a key concept to model the non-equilibrium experiments and develop protocols for the ultrafast orbital manipulation. In order to address the role of correlations in multi-band systems, we start from the simplest case of a two-band insulator in which the correlated half-filled valence band ($v$) and the conduction band ($c$) of different orbital origin are separated by a bare gap 2$\Delta_{\mathrm{bare}}$ (see Fig. \ref{fig:multiband}). The presence of strong onsite Coulomb repulsion contributes to the narrowing of the two bandwidths and makes the existing bare gap more robust. The effective gap thus takes into account that the $v$-band, corresponding to the LHB in the single-band Mott insulator, is pushed down in energy by a factor $U$/2. The effective band distance, 2$\Delta_{\mathrm{eff}}$, can be thus parametrized by the average occupation of each band \cite{Sandri2015}:     
\begin{equation}
\label{eff_gap}
2\Delta_{\mathrm{eff}}=2\Delta_{\mathrm{bare}}+\frac{U}{2}(n_v-n_c)=2\Delta_{\mathrm{bare}}+\frac{U}{2}p
\end{equation}
where $p$ represents the band (orbital) polarization, defined as the difference between the average occupations of the valence ($n_{v}$) and conduction ($n_{c}$) bands. The fully polarized case, i.e. $p$=1, corresponds to a Mott insulator in which the minimum effective gap is the effective distance between the LHB and the conduction band. When $U$ decreases, the simultaneous broadening of the Hubbard bands and the decrease of 2$\Delta_{\mathrm{eff}}$ drive a highly non-linear evolution of $p$. If 2$\Delta_{\mathrm{bare}}$ is of the order (or smaller) of $W$, the system eventually collapses into a two-band metallic state \cite{Sandri2015} in which the population of the two bands is the same, i.e. $p$=0.

It is evident from Eq. \ref{eff_gap} that the band polarization plays a crucial role in determining the electronic properties of multi-band correlated materials. At the same time, $p$ constitutes a new control parameter that can be manipulated by ultrashort light pulses. The possibility to transiently modify the charge distribution within bands of different orbital nature provides a novel non-thermal way to manipulate the electronic phases in multi-band TM compounds. Starting from the insulating state, the strong resonant excitation of specific interband transitions can lead to the collapse of the insulating gap and the formation of transient metallic phases. In the low excitation regime, the same excitation scheme can be used to probe the possible dependence of the bandstructure on the band (orbital) population, thus unveiling the underlying Mottness of the system \cite{Peli2017}. The other way around, a metallic state can be in principle driven into an insulating phase by proper tuning of the orbital occupation via ultrafast light stimulation.

In the following sections, we will discuss the role of the electronic correlations and of the presence of multiple bands (orbitals) in determining the main properties of some prototypical transition-metal compounds: superconducting copper oxides, Mott-insulating vanadium oxides and iron-based superconductors. These systems constitute the most promising platform for developing ultrafast orbital manipulation protocols. Finally, we will address some relevant properties of alkali-doped fullerides. Although fullerides do not belong to any TM compound class, they constitute another interesting example of a multi-band material on the verge of the Mott transition. These characteristics can be exploited to transiently manipulate the electronic distribution, up to the point of achieving a transient effective cooling of the quasiparticles \cite{Nava2017} upon ultrafast light excitation.  

\subsection{Emergent Mottness in high-$T_c$ superconducting copper oxides}
\label{sec:cuprates}
Copper oxides (cuprates) represent one of the most celebrated material in which electronic correlations drive the emergence of intriguing phenomena, culminating in high-temperature superconductivity exhibited by some families of cuprates where a suitable amount of charge carriers is introduced by means of chemical doping.
The key actors in determining the electronic properties of cuprates are the Cu 3-$d$ orbitals, filled by 9 electrons and subject to the local crystal field associated with the octahedral oxygen cage surrounding the Cu atoms. The 3-$d_{x^2-y^2}$ orbital plays a particularly important role since it is raised at the highest energy by the crystal field and hosts a single electron, thus behaving like a strongly-correlated half-filled band. The strong hybridization of the 3-$d_{x^2-y^2}$ levels with the ligand O-2$p_{x,y}$ orbitals calls for a multi-band description of the material. The ligand O-2$p_{x,y}$ levels form the O-2$p_{\sigma}$ band, which has a maximum at momentum (0,0), corresponding to a binding energy of $\sim$1.2 eV. In contrast, the O-2$p_{\pi}$ band, which arises from the hybridization of the O-2$p_{x,y}$ orbitals perpendicular to the O-2$p$ ligand orbitals, features almost no overlap with the conduction band and shows a maximum at momentum ($\pi$,$\pi$) and $\sim$1.2 eV binding energy.  

In the insulating parent compound, the half filled 3-$d_{x^2-y^2}$ band is completely suppressed by the strong local correlations, giving rise to the lower and upper Hubbard bands up to $\sim$10 eV far apart. Nonetheless, the minimum gap of the system is on the order of 1.5-2 eV and corresponds to the energy necessary to move an electron from the O-2$p_{x,y}$ to the Cu-3$d$ states, thus creating a double occupation on the 3-$d_{x^2-y^2}$ orbital. This process is usually referred to as charge-transfer (CT) excitation.
 
The concept of orbital polarization is a useful tool to model and interpret time-resolved experiments in cuprates. For sake of clarity, we will consider the antiferromagnetic insulating compound, where the CuO$_2$ unit constitutes the building block for the in-plane bandstructure. If we assume to sit on a copper spin-up site, the LHB will correspond to the energy of the $d_{\uparrow}$ electron, while the UHB to the energy necessary to add a $d_{\downarrow}$ electron to the same site. The oxygen bands, located between the LHB and UHB, provide the charge reservoir for populating the UHB. The minimum gap thus corresponds to the transition of a spin-down electron from the oxygen to the UHB. At equilibrium, and for very high values of U, the average occupation of the UHB, $\left\langle n_{\mathrm{UHB}}\right\rangle $, is almost zero while the average occupation of the three $p$ orbitals of each inequivalent oxygen (labeled by $i$=1,2) is $\left\langle n_{\mathrm{O}_{i,\sigma}} \right\rangle$=3 for each spin orientation (labeled by $\sigma$=$\uparrow$,$\downarrow$).
We can thus introduce the following orbital polarization:
\begin{equation}
p=\frac{\sum_{i,\sigma}\left\langle n_{\mathrm{O}_{i,\sigma}} \right\rangle- \left\langle n_{\mathrm{UHB}} \right\rangle-10}{2}
\end{equation}
which is equal to unity at equilibrium ($\sum_{i,\sigma}\left\langle n_{\mathrm{O}_{i,\sigma}} \right\rangle$=12; $\left\langle n_{\mathrm{UHB}} \right\rangle$=0) and progressively decreases as electrons are moved from the O to the Cu orbitals, down to the value $p$=0 when the UHB is completely filled ($\sum_{i,\sigma}\left\langle n_{\mathrm{O}_{i,\sigma}} \right\rangle$=11; $\left\langle n_{\mathrm{UHB}} \right\rangle$=1). The mean values of the Cu and O levels, i.e. $\mu_{\mathrm{Cu},\downarrow}$ and $\mu_{\mathrm{O}_{i,\sigma}}$, can be calculated via a simple mean-field model \cite{Peli2017} that provides the expression of the effective charge transfer gap as a function of the band polarization:
\begin{equation}
\label{CT_shift}
2\Delta^{\mathrm{CT}}_{\mathrm{eff}}=\mu_{\mathrm{Cu},\downarrow}-\mu_{\mathrm{O}_{i,\sigma}}= 2\Delta^{\mathrm{CT}}_{\mathrm{bare}}+\left(2U_{\mathrm{pd}}-\frac{5}{24}U_{\mathrm{pp}} \right) p
\end{equation}
where $U_{\mathrm{pp}}$ is the Coulomb repulsion between two charges occupying the same O-2$p$ orbital and $U_{\mathrm{pd}}$ is the Coulomb interatomic potential between the excess Cu-3$d$ electrons and the holes residing on the nearest neighboring oxygen sites.

\begin{figure} [t]
   \begin{center}
   \begin{tabular}{c}
   \includegraphics[height=13cm]{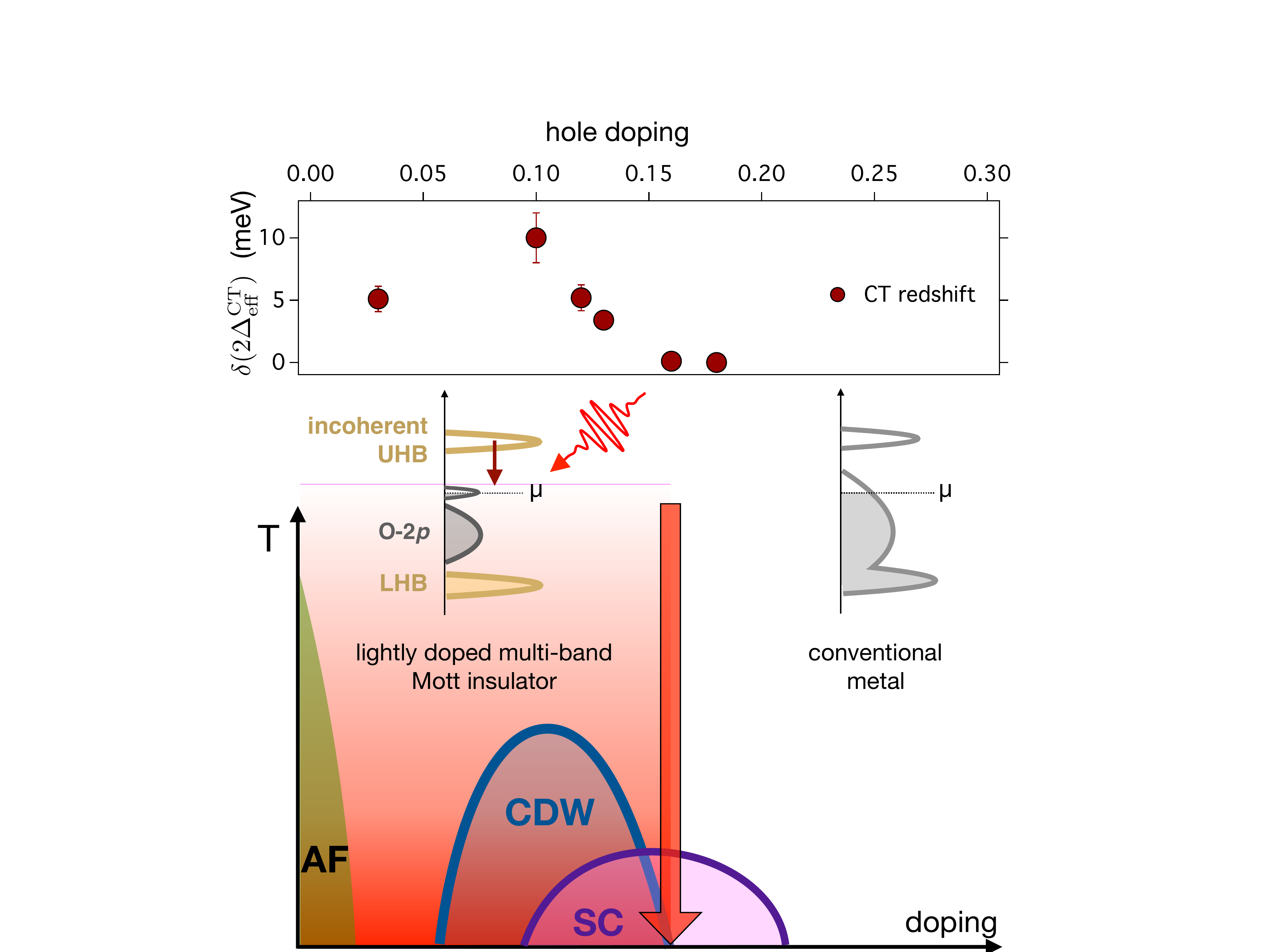}
   \end{tabular}
   \end{center}
   \caption[figure1] 
   { \label{fig_cuprates} 
Mottness in the phase diagram of superconducting copper oxides. Top panel: absolute value of the photoinduced CT shift for different hole doping concentrations. Bottom panel: sketch of the temperature-doping phase diagram. CDW: charge-density-wave region; AF: antiferromagnetic insulating state; SC: superconducting dome. Adapted from Ref. 	\citenum{Peli2017}}
   \end{figure} 
   
When chemically doped, the redistribution of the charges within the Cu-O orbitals and the consequent change of $p$ is at the base of many unconventional phenomena. In the lightly doped region of the phase diagram, the additional holes are preferably localized on the O orbitals. However, although metallic, lightly doped cuprates are characterized by the partial freezing of the electrons moving along the Cu-O bonds, which leads to the suppression of quasiparticles \cite{Norman1998} in the antinodal region of the Brillouin zone at momenta ($\pm\pi$,0) and (0,$\pm\pi$). This breakdown of the antinodal Fermi surface has been suggested \cite{Civelli2005,Ferrero2009,Rice2012,Alloul2014} to be the leading mechanism for the emergence of the pseudogap phase \cite{Timusk1999,Norman2005}, low-temperature charge instabilities \cite{Fradkin2012,Keimer2015} and $d$-wave superconductivity \cite{Tsuei2000} which dominate the low doping ($p<$0.16) region of the cuprate phase diagram.
It is now evident that the ultrafast resonant excitation of the CT transition can be used as an ``unconventional" control parameter \cite{Orenstein2012,Giannetti2016} to manipulate the orbital polarization and drive the system towards a transient non-thermal state in which the orbital occupation is almost reversed and closer to that of an uncorrelated metallic phase. This technique can be used to tackle some of the most important issues that still lack a definitive answer \cite{Keimer2015}:
\begin{itemize}
\item[i)] The way the lightly doped CT correlated insulator evolves into a conventional and less correlated metal as the hole doping increases
\item[ii)] The origin of the ubiquitous antinodal suppression of states in the underdoped region of the cuprate phase diagram
\item[iii)] The mechanism underlying the universal instability towards charge-ordered states in lightly doped cuprates
\end{itemize}

In ref. \citenum{Peli2017}, orbital manipulation has been used to investigate the high-energy Mottness in a prototypical cuprate family (Bi$_2$Sr$_{2-x}$La$_x$CuO$_{6+\delta}$, La-Bi2201) across a very broad range of hole dopings. The main idea driving the experiment is the possibility of modifying $p$ through the impulsive CT photoexcitation and measuring the transient redshift of $2\Delta^{\mathrm{CT}}_{\mathrm{eff}}$, as predicted by Eq. \ref{CT_shift}. Considering the fluence of the absorbed 10 fs-pulses at 1.35 eV photon energy (0.3 eV bandwidth), the orbital polarization variation is estimated as $\delta p\sim$ 0.3\%. Starting from realistic values of $U_{\mathrm{pp}}\sim$5 eV and $U_{\mathrm{pd}}\sim$2 eV \cite{Hansmann2014}, the estimated photoinduced CT redshift is $\delta(2\Delta^{\mathrm{CT}}_{\mathrm{eff}})$=$(2U_{\mathrm{pd}}-5/24U_{\mathrm{pp}})\delta p \simeq$-9 meV. Indeed a transient CT redshift has been observed by time-resolved optical spectroscopy in La-Bi2201 for low doping concentrations and at temperatures as high as 300 K (see Fig. \ref{fig_cuprates}). Interestingly, the photoinduced   $\delta(2\Delta^{\mathrm{CT}}_{\mathrm{eff}})$ renormalization was found to progressively vanish as the hole doping concentration was increased. For hole concentrations larger than the critical value $p_{cr}\simeq0.16$\%, corresponding to the doping necessary to attain the highest critical temperature ($T_c\sim$30 K) of the compound (optimal doping), no CT redshift was observed. This finding indicates the transition towards a more conventional metallic state with no reminiscence of Mott physics at the CT energy scale (see Fig. \ref{fig_cuprates}). 

As discussed in Sec. \ref{sec:intro}, the counterpart of the photoinduced high-energy modification of the Hubbard bands in correlated materials is the emergence of new conductive quasiparticle states close to the Fermi level. In the case of copper oxides the onsite Coulomb repulsion leads to the selective freezing of charges moving along specific $\textbf{k}$-directions, namely the momenta ($\pm\pi$,0) and (0,$\pm\pi$), corresponding to the antinodal region of the Fermi surface. In order to explore the dynamics of the antinodal quasiparticle states, a time-resolved extreme-ultra-violet (EUV) photoemission experiment has been recently performed \cite{Cilento2017}. The key feature was the use of ultrashort light pulses ($<$50 fs) with photon energy (18 eV) large enough to probe the antinodal region of the Brillouin zone. Near-infrared (1.65 eV) pump pulses were employed to resonantly excite the O-2$p_{\sigma,\pi}\rightarrow$Cu-3$d$ charge-transfer process in Bi$_2$Sr$_2$Ca$_{0.92}$Y$_{0.08}$Cu$_2$O$_{8+\delta}$ single crystals (Y-Bi2212) close to the optimal hole concentration ($T_c$=96 K). The pump excitation corresponded to the transfer of a fraction $\delta n_{\mathrm{UHB}}$=-($\sum_{i,\sigma}\delta n_{\mathrm{O}_{i,\sigma}}$)=$\delta p \simeq$1 \% of charges from the oxygen to the copper orbitals. After inducing such a non-thermal orbital polarization, the experiment unveiled the emergence of new transient quasiparticle states at the antinodes, while the nodal quasi-particle distribution was heated up as in a conventional metal. This observation directly confirmed the photoinduced antinodal metallicity previously argued by time-resolved optical spectroscopy \cite{Cilento2014}. As a further confirm of the intertwining of the oxygen and copper bands, the same experiment showed that the dynamics of the transient antinodal metallicity is mapped into the dynamics of the O-2$p_{\pi}$ bands \cite{Cilento2017}, thus directly demonstrating the importance of the multiband description to model the non-equilibrium physics of copper oxides.

The use of ultrafast excitation to manipulate the orbital polarization shed new light on some important aspects of the physics of superconducting copper oxides:
\begin{itemize}
\item[-] time-resolved optical spectroscopy demonstrated that the phase diagram of copper oxides is dominated by a temperature-independent transition at $p_{cr}\simeq$0.16 from a correlated system, whose properties at the CT energy scale are similar to those of a multi-band Mott insulator, to a more conventional metal, 
\item[-] the suppression of quasiparticle states at the antinodes is originated by a correlation-driven breakup of the Fermi surface. The optical manipulation of $p$ drives the evolution of antinodal states from Mott-like gapped excitations to delocalized quasiparticles, 
\item[-] finally, the outcomes of the time-resolved spectroscopies, summarized in the previous two points, suggest that the universal emergence of low-temperature charge instabilities is the low-energy consequence of the Mottness that characterizes the $p<p_{cr}$ region of the phase diagram \cite{Peli2017}. 
\end{itemize}

                                                                                                                                                                                                                                                                                                                                                                                                      \subsection{Photoinduced insulator-to-metal transition in V$_2$O$_3$}
\label{sec:vanadates}

   \begin{figure} [t]
   \begin{center}
   \begin{tabular}{c} 
   \includegraphics[height=11cm]{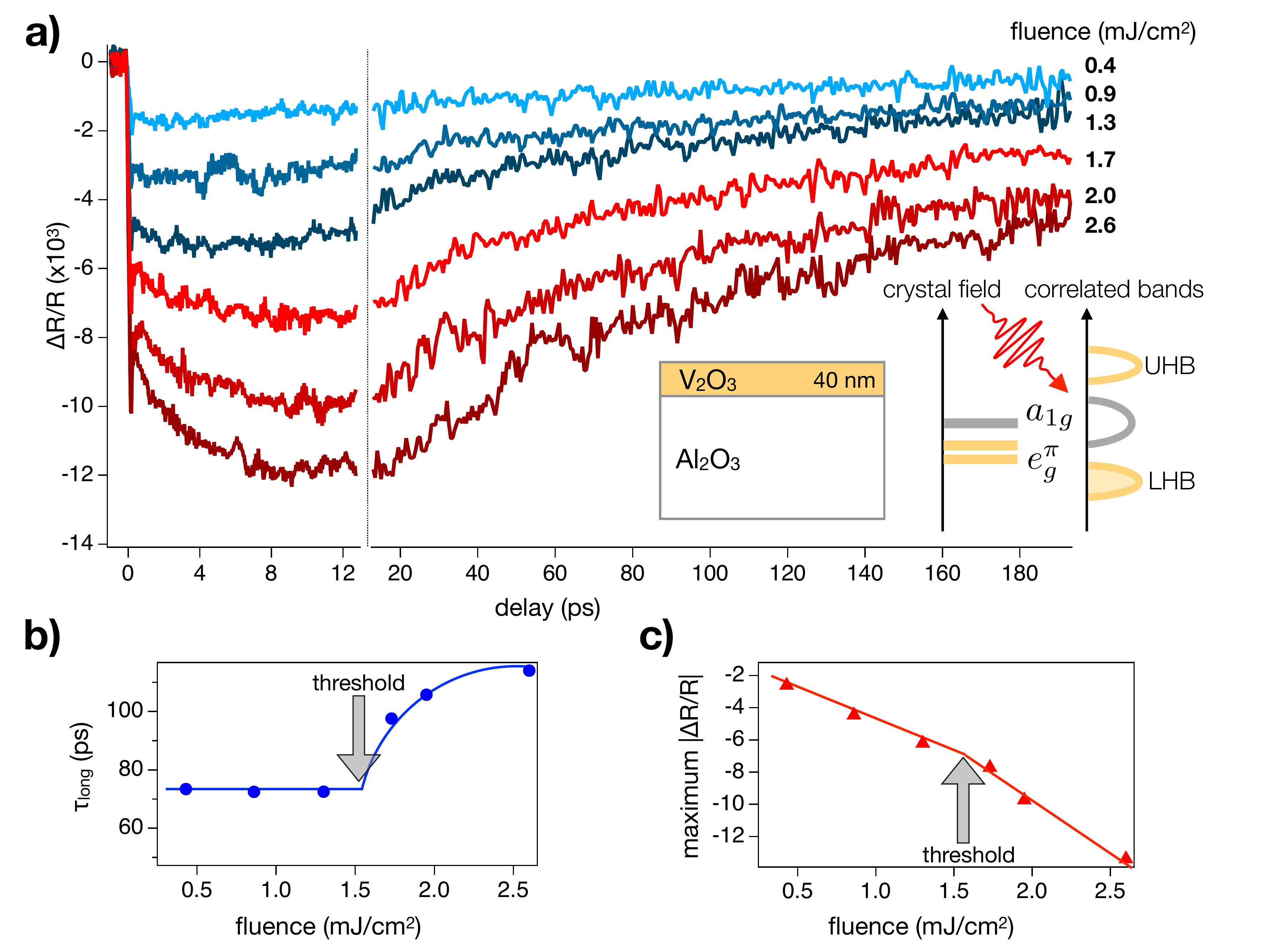}
   \end{tabular}
   \end{center}
   \caption[figure1] 
   { \label{fig_V2O3} 
Photoexcitation of the insulator-to-metal phase transition in V$_2$O$_3$. \textbf{a)} Time-resolved reflectivity variations at $T$=80 K and for different pump fluences. The pump photon energy (1.5 eV) is resonant with the transition from the LHB to the $a_{1g}$ band. The probe photon energy is $\sim$2 eV. A sketch of the sample structure and of the V$_2$O$_3$ electronic levels is shown in the bottom-right panel. \textbf{b)} Pump-fluence dependence of the slow relaxation timescale ($\tau_{\mathrm{long}}$). The solid line is a guide to the eye. \textbf{c)} Pump-fluence dependence of the maximum reflectivity variation. The solid line is a guide to the eye.}
   \end{figure}

Vanadium oxides are among the most extensively studied correlated materials for the richness of their phase diagrams. In these compounds, the interplay between the strong on-site correlations on the V-3$d$ orbitals and the lattice-driven lift of their degeneracy gives rise to spectacular insulator-to-metal phase transitions \cite{Imada1998}. In VO$_2$ and V$_{2}$O$_3$ compounds the electrical resistivity \cite{Kuwamoto1980} can be tuned across several order of magnitude by means of small variations of chemical composition \cite{McWhan1969,Imada1998,Lupi2010,Homm2015,delValle2017}, pressure \cite{Limelette2003,Jayaraman1970,Rodolakis2010,Valmianski2017}, temperature \cite{Qazilbash2007,Stewart2012,McLeod2017}, and, more recently, by external application of electric fields \cite{Kim2010,Nakano2012,Guenon2013,Stoliar2013,Mazza2016} or irradiation with short light pulses \cite{Cavalleri2001,Liu2011,Mansart2010,Giannetti2016,Lantz2017,Basov2017}. Here, we will focus on vanadium sequioxide (V$_2$O$_3$), that is considered a paradigmatic example of multi-band Mott insulator. V$_2$O$_3$ exhibits an electronic transition from a low-temperature antiferromagnetic insulating phase ($\sim$0.7 eV energy gap) to a high-temperature paramagnetic metallic state at $T_{IMT}\sim$ 160 K. Furthermore, V$_2$O$_3$ presents different crystal structures in the two phases: the monoclinic unit cell of the antiferromagnetic insulating phase turns into a rhombohedral unit cell in the paramagnetic metallic phase. 

The electronic bandstructure of vanadium sesquioxide strongly depends on the local crystal field in which the V-3$d$ orbitals are immersed. Starting from the [Ar]3$d^3$4$s^2$ atomic configuration, the triple oxidation leaves two electrons in the V-3$d$ levels. In the insulating monoclinic phase, the octahedral crystal-field splitting yields to the separation of 3$d$ orbitals into a lower $t_{2g}$ and an upper $e_g$ band. Since the octahedra show further trigonal distortion, reinforced by the on-site Coulomb repulsion, the $t_{2g}$ are split into a lower doubly degenerate $e^{\pi}_{g}$ and a single bonding $a_{1g}$ state which is pushed at an energy of about 2$\Delta_{\mathrm{eff}} \sim$1.5 eV above the $e^{\pi}_{g}$ levels \cite{Poteryaev2007,Grieger2015}.
Since in the ground state the occupation of the $e^{\pi}_{g}$ levels is almost one, the system behaves as a half-filled two-band Mott insulator. At $T_{IMT}$, the structural change from monoclininc to corundum accompanies a jump of the $a_{1g}$ occupation \cite{Park2000} and the consequent transition to the metallic phase. The band polarization, defined as $p$=$n_{e^{\pi}_{g}}$-$n_{a_{1g}}$, has been suggested \cite{Sandri2015,Lantz2017} to be the control parameter for the insulator-to-metal phase transition and for the effective gap between the occupied and $e^{\pi}_{g}$ and the empty $a_{1g}$ levels.

Optical pumping can be used as a direct tool to manipulate the band polarization in V$_2$O$_3$ and eventually, inducing the collapse of the insulating gap. A sudden transfer of about $\delta p \simeq$0.1 electrons/site from the LHB to the $a_{1g}$ conduction band has been predicted to be the threshold to make the antiferromagnetic insulating phase unstable and drive the system into a metastable transient metallic state \cite{Sandri2015,Lantz2017}. In Fig. \ref{fig_V2O3} we report the outcomes of time-resolved optical spectroscopy on 40 nm V$_2$O$_3$ thin films epitaxially grown on a Al$_2$O$_3$ substrate. The pump photon energy (1.5 eV) drives direct transitions from the V-3$d$ LHB to the $a_{1g}$ band. The dynamics of the relative reflectivity variation, defined as the normalized difference between the excited and equilibrium reflectivities ($\Delta R/R$=($R_{\mathrm{exc}}$-$R_{\mathrm{eq}}$)/$R_{\mathrm{eq}}$), is recorded at $\hbar\omega_{\mathrm{probe}}$=2 eV for different pump excitation fluences. The sample temperature is $T$=80 K, well below the equilibrium $T_{IMT}$.  At low fluences, the sudden $\Delta R/R$ signal monotonically decreases as an exponential decay characterized by the timescale $\tau_{\mathrm{long}}\simeq$70 ps. Above a threshold fluence ($F_{\mathrm{thr}}$) of $\sim$1.5 mJ/cm$^2$, an additional build-up dynamics is clearly observed in the 0-15 ps time window. This timescale roughly corresponds to the ultrafast photoinduced change of symmetry recently observed via time-resolved X-ray scattering \cite{Singer2017}.  At the same time, the relaxation dynamics described by $\tau_{\mathrm{long}}$ is significantly slowed down as compared to the low fluence data. In Fig. \ref{fig_V2O3}b) and c) we report the fluence dependence of $\tau_{\mathrm{long}}$ and of the maximum $|\Delta R/R|$ signal. A clear transition is observed at $F_{\mathrm{thr}}\sim$1.5 mJ/cm$^2$, thus suggesting that the photoinduced insulator-to-metal transition is taking place. This value is very close to the reported threshold value measured by near infrared pump-THz probe \cite{Liu2011}. Considering the light absorption coefficient ($\alpha$=0.58$\cdot$10$^5$ cm$^{-1}$, Refs. \citenum{Baldassarre2008,Qazilbash2008}) at $\hbar\omega_{\mathrm{pump}}$=1.5 eV, it is possible to estimate the energy density absorbed by the thin film, $E_{\mathrm{abs}}$=80 J/cm$^3$, which corresponds to the photon density $n_{\mathrm{ph}}$=3.3$\cdot$10$^{20}$ cm$^{-3}$. This density approximately corresponds to 0.1 photon/unit cell or 0.01 photon/V atom. The discrepancy between the experimental orbital polarization photoinduced at the threshold fluence ($\delta p\simeq$0.01) and the estimated value for achieving an impulsive collapse of the insulating phase ($\sim$0.1) calls for the consideration of photothermal effects and cooperative lattice phenomena in the model.

\subsection{Manipulation of orbital-selective Mottness in iron-based superconductors}
\begin{figure} [t]
   \begin{center}
   \begin{tabular}{c}
   \includegraphics[height=10cm]{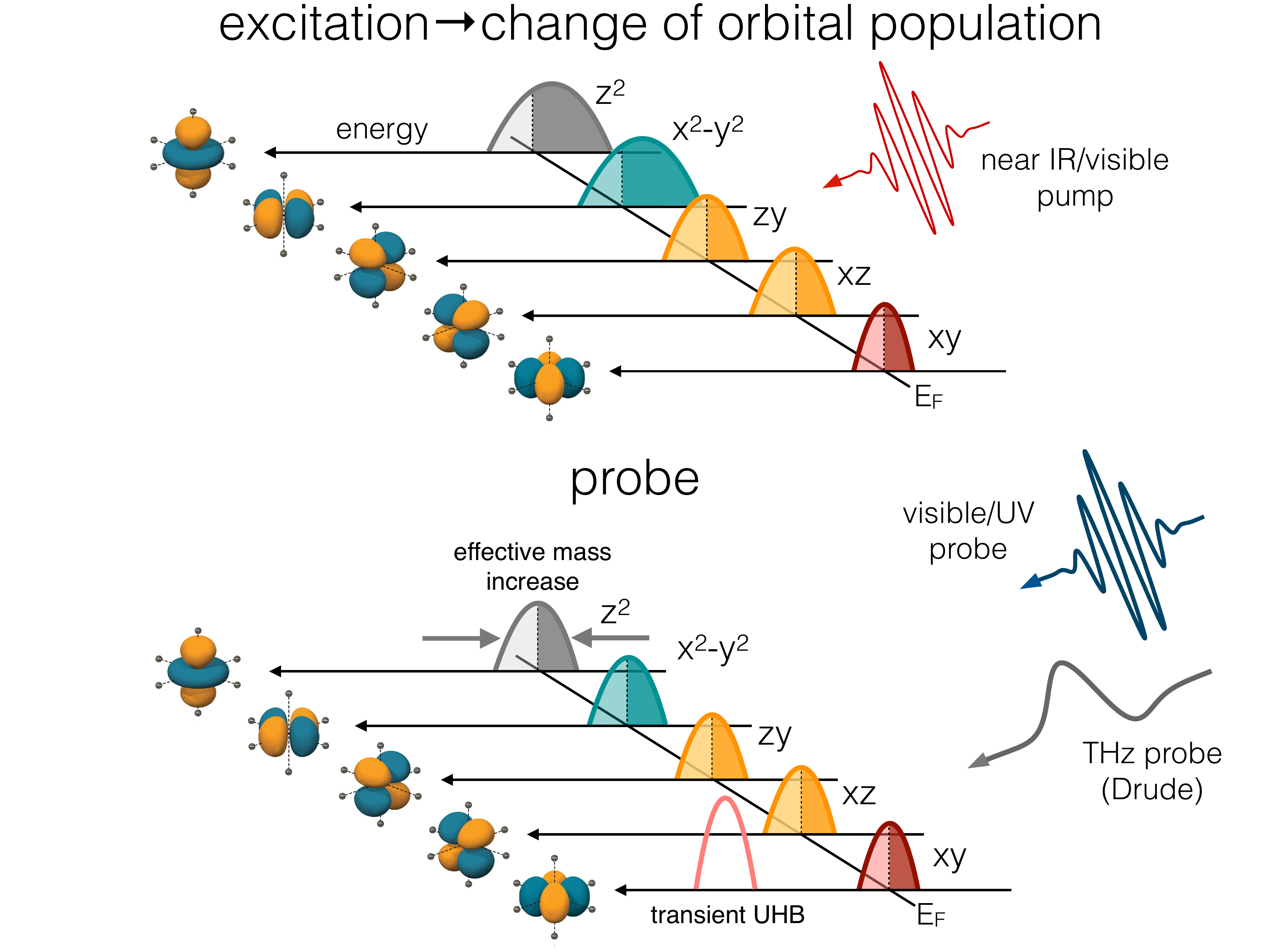}
   \end{tabular}
   \end{center}
   \caption[figure1] 
   { \label{figure1} 
Cartoon of the ultrafast orbital manipulation in iron-based superconductors. The selective change of occupation of the d$_{xy}$ orbitals, which are the closest to the Mott insulating state, can drive the emergence of a transient orbital-specific Mottness characterized by the quench of metallic states at the Fermi level and the appearance of high-energy Hubbard bands.}
   \end{figure} 
   The multiband character of the electronic structure is perhaps the most distinctive feature of another family of correlated superconductors, the iron-based superconductors. In this section we discuss how the combination of the electronic interactions and the multi-orbital nature of the bands leads to novel opportunities to control the electronic properties of these materials by properly tuning the occupation of different orbitals.

One of the most important open questions about iron-based superconductors is the degree of similarity between them and the cuprates, which includes the question about the role of electronic correlations. From a structural point of view, the members of the different families share indeed a common building block, layers of iron atoms and a ligand which can be a pnictogen (As) or a chalcogen (Se, Te) atom. These layers somehow replace the copper-oxygen layers with some important differences in the electronic structure. The generic phase diagram also shares similarities with the cuprates as superconductivity generally appears when doping a spin-density wave metal\cite{Nature_paglione}. 

The replacement of the antiferromagnetic charge-transfer insulator of the cuprates with a metal with magnetic ordering supports, on one hand, a magnetic mechanism for superconductivity\cite{Mazin}, but on the other hand seems to imply a weaker degree of correlation with respect to the cuprates. This has led the community to argue for an intermediate degree of correlation and to develop theoretical approaches based either on a weak- or on a strong-coupling picture \cite{Si2013,Chubukov2015,DeMedici2015}. However, in the last years it emerged more and more clearly that these assumptions hides a crucial feature of these materials, namely the coexistence of electrons with a completely different degree of correlation\cite{deMedici2009}.

A key element to understand the ``selective" correlations in iron-based superconductors is the multiorbital character of the electronic structure. It is universally recognized that the Fermi surface of these materials is formed by several pockets and that the description of the relevant fermionic degrees of freedom requires to include all of the five Fe-$d$ orbitals, while the inclusion of the ligand $p$ orbitals appears less important \cite{Graser2009}.  

The stoichiometric compounds correspond to a filling of 6 electrons per Fe atom in the five bands derived by iron $d$ orbitals. The filling is thus integer, which would be compatible with a Mott transition, but it is different from global half-filling, which would require 5 electrons per atom. Another crucial observation is a sizable value of the Hund's coupling exchange, which favours high-spin configurations and it has been found to strongly affect the strength and the nature of the electronic correlations \cite{1367-2630-11-2-025021}. 
These observation led indeed to a surge of studies of electronic correlations in multiorbital Hubbard models, which have clarified a number of open issues. We refer to Refs. \citenum{GeorgesAR2013,Bascones2016,deMedici2017} for reviews. 

As far as the iron-based materials are concerned, it is crucial to note that the Hund's coupling strongly reduces the critical interaction strength $U_c$ for a Mott transition at global half-filling (n=5) while it increases $U_c$ at $n=6$. This is the reason why the parent compounds are not Mott insulators. On the other hand, it has been shown that, in the relevant regime for iron-based materials, the Hund's coupling induces remarkable correlations, as measured by the effective mass. Furthermore, it has been shown that the Hund's coupling ``decouples" the different orbitals so that an excitation created in a given orbital remains with the same orbital index with very small orbital fluctuations.  This is a direct consequence of the selection of a high-spin state, which implies to spread the electrons as much as possibile among the orbitals. The combination of the Hund's coupling and the multi-orbital nature of the bands leads to two main consequences: 

\begin{itemize}
\item{the degree of correlation increases as the number of electrons is reduced by hole doping and decreases with electron doping,}
\item{the increase of the correlations is strongly orbital-selective. Namely electrons belonging to different orbitals can have sizably different effective mass.}
\end{itemize}

These two observations, which are based on experimental observations alone, can be rationalized in a very simple and nice picture. The degree of correlation is indeed controlled by the distance in doping from the Mott insulator at $n=5$ which would be realized if we could dope the sufficient number of holes\cite{PhysRevLett.112.177001}. However, the orbitals are decoupled by the Hund's term\cite{PhysRevLett.112.177001,PhysRevB.92.075136}, which implies that each individual orbital has a distinct degree of correlation depending on its own population and, in particular, on its distance from half-filling. As a matter of fact, orbitals which have an occupation closer to $n=1$ are more correlated than those having a larger occupation. This concept remarkably explains the above described phenomenology and points out that orbital-selective correlations are a key to understand the nature of the correlated metallic state of iron-based materials. Therefore, even if the parent compounds are not Mott insulators, they can still be seen as doped Mott insulators, where the reference is the Mott insulator which would be realized at $n=5$. This picture has been confirmed by measurments of the Sommerfeld coefficient\cite{PhysRevB.94.205113} and x-ray spectroscopy\cite{PhysRevB.96.045133}.

This succesful theoretical scheme has huge implications for the orbital control of the electronic properties of these materials. In particular if we could selectively control the population of some orbitals, we would be able to tune their conduction properties making them more metallic if the occupation is driven far from half-filling, or even insulating if the population is tuned at half-filling. There are indeed members of the family where some orbitals are actually half-filled and are therefore insulating\cite{PhysRevB.91.085124} and we can turn them metallic with a similar selective excitation. 

The possibility to switch on and off the conduction of selective channels can be used to obtain complex conduction patterns exploiting the different shape of the orbitals and also to control the other electronic properties of the iron-based superconductors, including obviously superconductivity and nematicity (the breaking of the lattice rotational symmetry in a metallic phase), whose connection with the physics of electronic correlations has been discussed in Ref. \citenum{PhysRevB.95.144511}. The actual implementation of protocols able to selectively excite different orbitals can thus shed light on the possible active role of electronic correlations in the superconducting mechanisms and possibly lead to optimized superconducting properties. A first interesting result in this direction has been reported in Ref. \citenum{DeNinno}, where an orbital-selective excitation process has been proposed and realized.

\subsection{Transient effective cooling in alkali-doped fullerides}
\label{sec:fullerides}
\begin{figure} [t]
   \begin{center}
   \begin{tabular}{c} 
   \includegraphics[height=15cm]{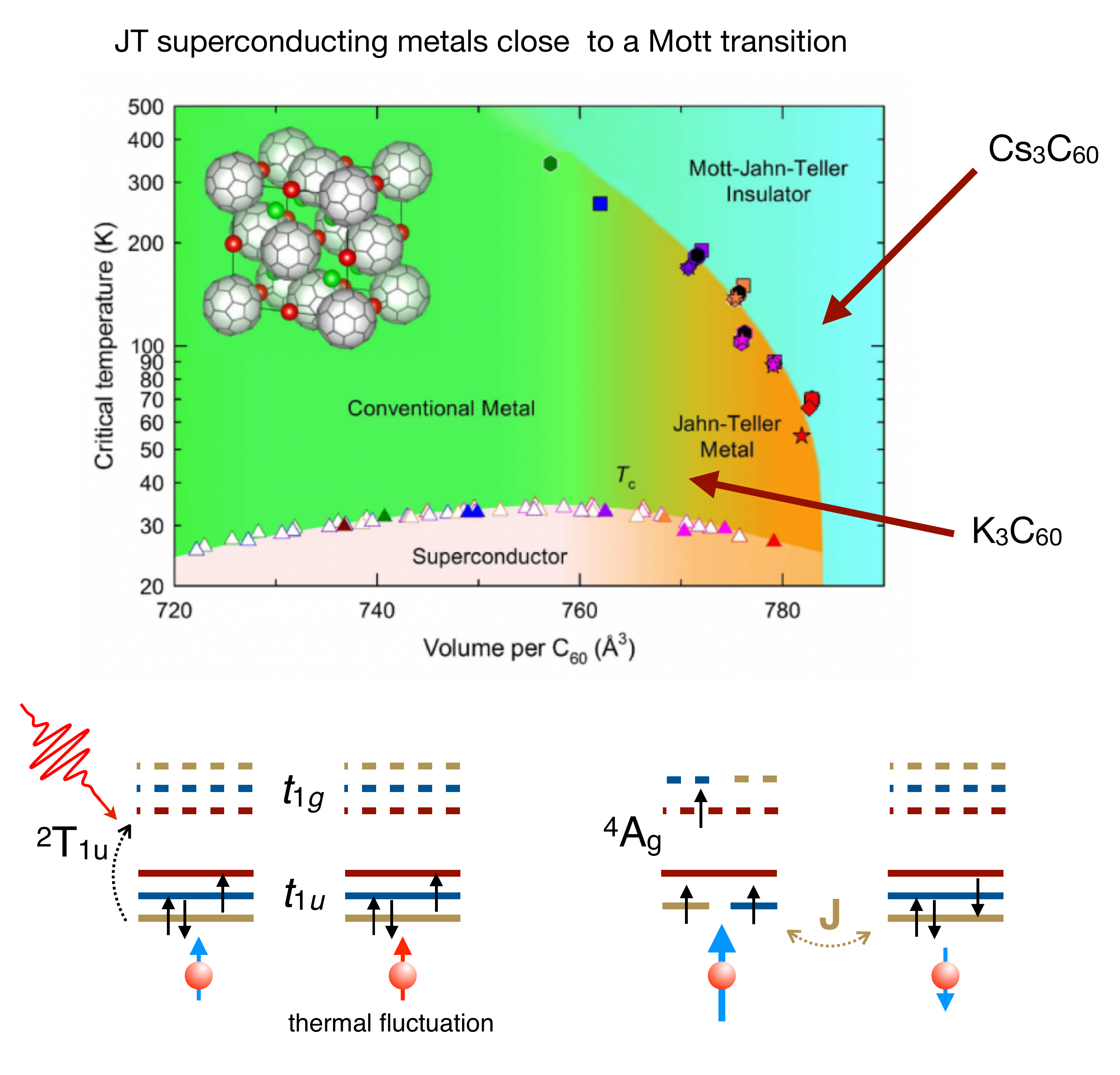}
   \end{tabular}
   \end{center}
   \caption[figure1] 
   { \label{fig_C60} 
Photoexcitation of the local triplet super-exciton in K$_3$C$_60$. Top panel: phase diagram of alkali-doped fullerides. K$_3$C$_60$ is close to the Mott-Jahn-Teller insulating phase and exhibits a superconducting phase at low temperature. Adapted from Ref. \citenum{Zadik2015}. Bottom panel: sketch of the photoexcitation process which drives the formation of long-lived local super-excitons through the absorption of a paramagnon.}
   \end{figure}

Alkali-doped fullerides, A$_3$C$_{60}$, represent an interesting class of correlated materials, in which the interplay between the on-site Coulomb interactions and the high icosahedral symmetry of C$_{60}$ favors a strong Jahn-Teller coupling between the electronic states and the lattice \cite{Gunnarsson1997}. A$_3$C$_{60}$ is a molecular crystal in which the crystal-field splits the carbon electronic levels into $t_{1u}$ lowest unoccupied molecular orbitals (LUMO) and $t_{1g}$ orbitals at higher energy. The threefold degenerate $t_{1u}$ LUMO host the three electrons donated by the alkali metals. The simultaneous effects of the Jahn-Teller distorsions, the Hund's rule exchange and the incipient Mott localization (see phase diagram in Fig. \ref{fig_C60}) make the low-spin state S=1/2 the molecular ground state and eventually drive $s$-wave superconductivity \cite{Capone2002,Capone2009,Nomurae1500568} at temperatures as high as 20 K for K$_3$C$_{60}$ and 38 K for Cs$_3$C$_{60}$ under high pressure ($\sim$7 kbar) \cite{Gunnarsson1997}.

Interestingly, the optical conductivity of K$_3$C$_{60}$ \cite{Degiorgi1994} presents a broad absorption peak in the mid-infrared (20-200 meV) whose origin remained unclear until recently. In Ref. \citenum{Nava2017} it has been argued that this electronic transition corresponds to the optical excitation of a localized S=3/2 ”super-exciton” involving the promotion of one electron from the $t_{1u}$ half-filled correlated band to a higher-energy empty $t_{1g}$ state. This singlet-triplet transition, broadened by Franck-Condon effects, is in principle dipole forbidden but becomes allowed via the simultaneous absorption of a paramagnon. Exploiting the long lifetime ($>$1 ps) of these excitons and the abundance of paramagnon excitations in the vicinity of the Mott insulating state, it is possible to photoexcite a large number of localized excitons, which act as a sink for the energy provided by mid-infrared pulses. 

The excitation of high-energy super-excitons at energy $E_{\mathrm{SE}}$ via mid-infrared pumping can be used to transiently manipulate the population ($n_{t_{1u}}$) of the $t_{1u}$ conduction band with fundamental impact on the transient electronic properties of A$_3$C$_{60}$. The most striking consequence of such a band manipulation is that the photoinduced variation of the $t_{1u}$ population is inherently accompanied by the creation of a non-thermal distribution of localized super-excitons, $\delta N_{\mathrm{SE}}$=-$\delta n_{t_{1u}}$, and the simultaneous annihilation of thermally excited paramagnons. This process leaves the $t_{1u}$ quasiparticles in a smaller-entropy state, which corresponds to a transient decrease of the initial effective temperature \cite{Nava2017}. The dynamics of the transient quasiparticle entropy variation ($\delta S_{\mathrm{QP}}$) can be described by an approximated transient entropy equation, derived under the assumption of instantaneous local equilibrium:
\begin{equation}
T(t)\delta\dot{S}_{\mathrm{QP}}(t)=-[\omega-E_{\mathrm{SE}}+\mu(t)]\delta n_{t_{1u}}
\end{equation}  
where $\omega$ is the frequency of the incoming light and $\mu(t)$ the instantaneous chemical potential. Considering realistic excitation fluences, a transient cooling of the order of several tens of degrees has been predicted \cite{Nava2017}. On a timescale of few picoseconds the excitons would eventually decay, thus leading to a steady-state increase of temperature. This mechanism has been proposed as the possible explanation of the recently claimed transient superconductivity, photoinduced by pulsed mid-infrared irradiation in K$_3$C$_60$ at temperatures as high as ten times the equilibrium $T_c$ \cite{Mitrano2016,Cantaluppi2017}.

\subsection{Towards coherent control in solids}
\label{sec:coherent_control}
As discussed in this manuscript, the possibility of manipulating the band population and transiently modifying the orbital polarization opens new intriguing routes to probe and control the complex properties of multi-band materials.
Here we have provided many different examples of new concepts emerging from the transient orbital manipulation in correlated materials: the discovery of Mott-like physics regulating the phase diagram of superconducting copper oxides at hole doping as high as 16\% and $T$=300 K; the manipulation of electronic phases via band reversal in vanadium oxides; the control of orbital selective correlations in iron-based superconductors and the photoinduced effective cooling in alkali-doped fullerides.

   \begin{figure} [t]
   \begin{center}
   \begin{tabular}{c} 
   \includegraphics[height=10cm]{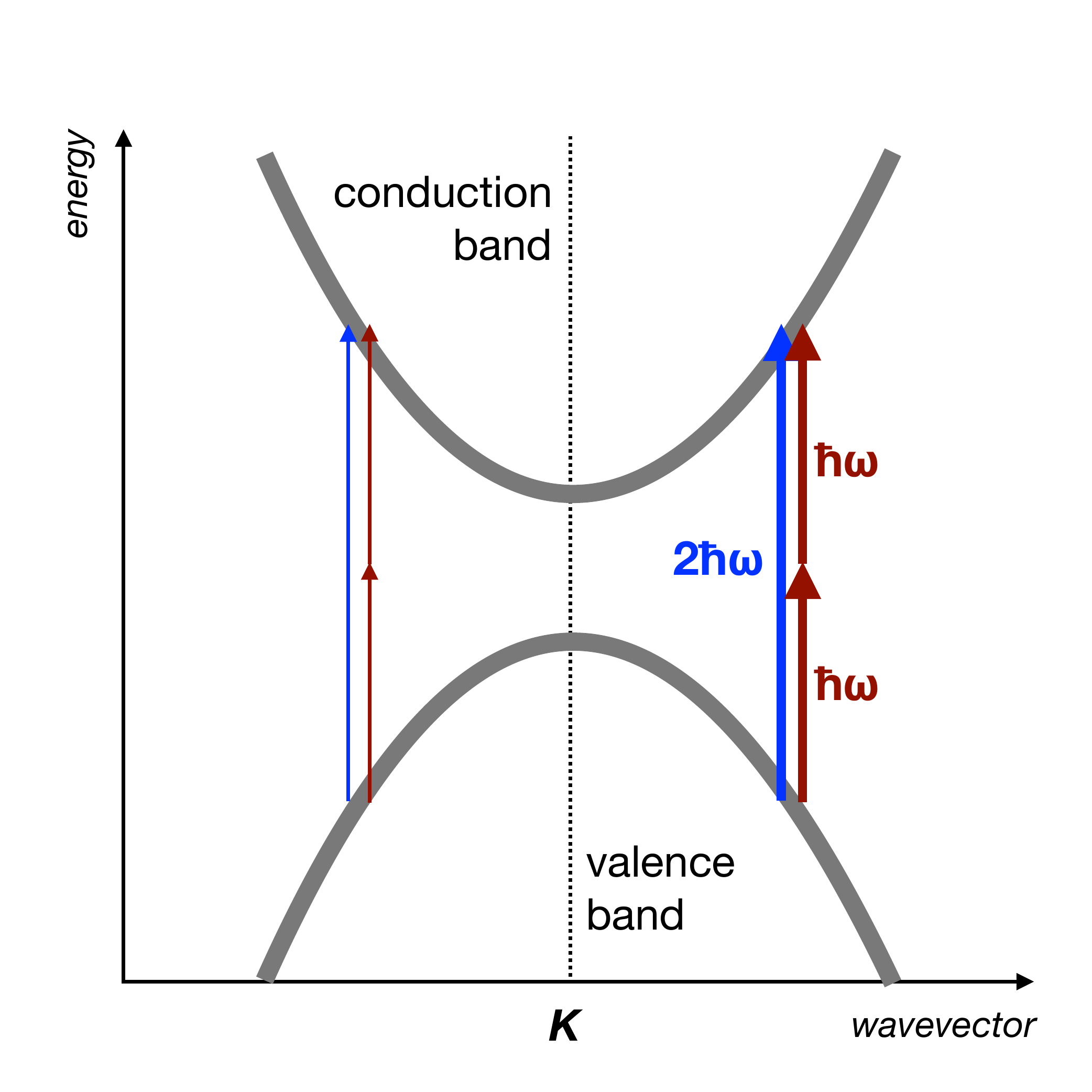}
   \end{tabular}
   \end{center}
   \caption[figure1] 
   { \label{fig_coherent_control} 
Cartoon of the coherent control protocol. The interference between the direct absorption process at 2$\hbar\omega$ and the two-photon process at $\hbar\omega$ gives rise to an excitation asymmetry at momenta {\it{\textbf{K}}}+{\wv} and {\it{\textbf{K}}}-{\wv}, where {\it{\textbf{K}}} is a generic high-symmetry wavevector of the crystal. The relative phase of the 2$\hbar\omega$ and $\hbar\omega$ fields regulates the direction of the excitation asymmetry.}
   \end{figure} 

Despite the recent impressive advances in the field and the many problems that still remain unexplored, the ``conventional" pump excitation scheme suffers from an intrinsic limitation, that is the lack of selectivity in the momentum-space. In simple terms, the pump excitation can be tuned across an optical resonance thus coupling to specific electronic transitions, independently of the \textbf{k} of the corresponding quasiparticle states. This limitation is particularly severe for multi-orbital materials in which the orbital component, along with the relative degree of correlation, of the valence and conduction bands can be strongly \textbf{k}-dependent. For example, the possibility of using ultrashort light pulses as a sort of optical ``tweezer" that selectively modifies the population of specific regions in the Brillouin zone, would allow to manipulate the correlations in multi-band materials in a momentum-resolved fashion. As a paradigmatic example, the use of light to unbalance the nodal-antinodal quasiparticle populations in superconducting copper oxides would allow to create new transient states with emergent properties with no counterpart at equilibrium. Similarly, the \textbf{k}-selective change of population in the conduction band of iron based superconductors would pave the way to the real control of the orbital polarization and to the possibility of turning a specific orbital component into a half-filled Mott insulating state and viceversa.

Among the different strategies to achieve the \textbf{k}-selective control of electronic excitations, one of the most promising is the exploitation of the quantum interference effect between two different quantum excitation pathways to affect the final state population. In the simplest configuration, this technique, named as \textit{coherent control}, is based on the use of two phase-locked beams at a fundamental ($\omega$) and second-harmonic (2$\omega$) frequencies that impinges on a crystalline material \cite{BookMeier}. The two equivalent pathways consist of two- and one-photon absorption processes of the fundamental and second-harmonic, which can interfere leading to an unbalanced creation of electron-hole excitations at momenta +\textbf{k} and -\textbf{k} with respect to high symmetry points of the Brillouin zone (see Fig. \ref{fig_coherent_control}) \cite{Sipe1996, Sipe2010, Sipe2011}.

In order to better understand the mechanism behind the  \textit{coherent control} protocol, it is instructive to calculate the carrier injection rate (or orbital polarization) in a simple two-band model ($v$: valence band; $c$: conduction band) by means of time-dependent perturbation theory. The states of interest are the ground (initial) state $\left| 0 \rangle \right. $ and states of the form $\left|  cv,\wv \rangle \right.=\ElectronCreation \HoleCreation \left| 0 \rangle \right.$, where $\ElectronCreation$ ($\HoleCreation $) creates an electron (hole) at wave vector $\wv$ in the conduction (valence), $c$ ($v$), bands. In the presence of a perturbation given by a classical electromagnetic field (which is composed by two fields at frequencies $\omega$ and $2\omega$), the state describing the system at time $t$ can be developed as $\left|  \Psi \! \left( t \right) \rangle \right.=c_0 \! \left( t \right) \left| 0 \rangle \right.  + c_{\scriptsize{cv, \wv}}\! \left( t \right) \left|  cv,\wv \rangle \right.$. Starting from the Fermi's golden rule, the orbital polarization change rate takes the form:
\begin{eqnarray} \label{CarrierDensityInjectionRate}
\delta\dot{p}(\wv)=\frac{2}{V}  \frac{\mbox{d}}{\mbox{d} t} \left| c_{\scriptsize{cv, \wv}} \right|^2=\frac{4\pi}{V} \, \left| K_{\scriptsize{cv, \wv}} \right|^2 \; \delta\Big( \omegacv -2\omega  \Big),
\end{eqnarray}
where $\delta p$=$\delta$($n_v$-$n_c$), $V$ is the normalization volume of the crystal and $\omegacv=\omegac-\omegav$, $\omegac$ and $\omegav$ representing the dispersion relations of the conduction and valence band, respectively. In this formulation, the matrix element $\left| K_{\scriptsize{cv, \wv}} \right|^2$ reads: 
\begin{eqnarray} \label{Kappacv}
\begin{aligned} 
\left| K_{\scriptsize{cv, \wv}} \right|^2=&\zeta_{1,cv}^{ij}\!\left( \wv \right) \, E^{i}\! \left( -2\omega \right) E^{j}\! \left( 2\omega \right)+\zeta_{2,cv}^{ijkl}\!\left( \wv \right) \, E^{i}\! \left( -\omega \right) E^{j}\! \left( -\omega \right) E^{k}\! \left( \omega \right) E^{l}\! \left( \omega \right)\\
&+\Big[\zeta_{3,cv}^{ijk}\!\left( \wv \right)\, E^{i}\! \left( -\omega \right) E^{j}\! \left( -\omega \right) E^{k}\! \left( 2\omega \right) +\mbox{c.c.} \Big],\\
\end{aligned}
\end{eqnarray}
where superscripts indicate Cartesian components (to be summed over if repeated), $E\! \left( \omega \right)$ and $E\! \left( 2\omega \right)$ are the electric field amplitudes at the indicated frequencies and $\hat{\zeta_{1}}$, $\hat{\zeta_{2}}$ and $\hat{\zeta_{3}}$ are second-, fourth-, and third-rank tensors. The three different terms of the right-hand side of Eq. \ref{Kappacv} represent: 1) the pure one-photon contribution; 2) the pure two-photon contribution; 3) the interference between the two-photon and one-photon transition processes. While $\hat{\zeta_{1}}$ and $\hat{\zeta_{2}}$ are even under the symmetry operation +\wv$\rightarrow$-{\wv} with respect to a high-symmetry wavevector, the $\hat{\zeta_{3}}$ is odd under the same operation and is thus responsible for the asymmetric distribution of the injected carriers around the high-symmetry point. 

This coherent excitation protocol has been recently used for  launching ballistic current pulses in arbitrary directions in different kinds of systems, ranging from molecules to solids, such as graphene \cite{BookMeier,Sipe2010,Sipe2011} and has been extended to the case of two different polarization states of light \cite{Sipe1999}. All these techniques constitute a promising new doorway to the optical manipulation of multi-band solids and the consequent control of the orbital polarization and electronic correlations in the {\wv}-space, with particular impact to the case of materials, such as cuprates and iron-based superconductors, in which the degree of correlation has a strong orbital and momentum dependence.

\section*{ACKNOWLEDGMENTS}
C.G. acknowledges support from Universit\`a Cattolica del Sacro Cuore through D1, D.2.2 and D.3.1 grants. M.C. and C.G. acknowledge financial support from MIUR through the PRIN 2015 program (Prot. 2015C5SEJJ001). This research was undertaken thanks in part to funding from the Max Planck-UBC Centre for Quantum Materials and the Canada First Research Excellence Fund, Quantum Materials and Future Technologies Program. The research activities of M.F. have received funding from the European Union, under the project ERC-692670 (FIRSTORM).
\bibliography{biblio_SPIE} 
\bibliographystyle{spiebib} 

\end{document}